# Performance Analysis of Energy Detection over Mixture Gamma based Fading Channels with Diversity Reception

Omar Alhussein*, Ahmed Al Hammadi†, Paschalis C. Sofotasios‡§, Sami Muhaidat*†,
Jie Liang*, Mahmoud Al-Qutayri†, and George K. Karagiannidis†§
*School of Engineering Science, Simon Fraser University, Burnaby, BC, Canada, V5A 1S6.
{oalhusse, jiel}@sfu.ca
†Khalifa University of Science Technology and Research, PO. BOX 127788, Abu Dhabi, UAE.
{100037703@kustar.ac.ae, muhaidat@ieee.org, mqutayri@kustar.ac.ae}
‡Department of Electronics and Communications Engineering, Tampere University of Technology, 33101 Tampere, Finland.
{p.sofotasios@ieee.org}
§Department of Electrical and Computer Engineering, Aristotle University of Thessaloniki, 54124 Thessaloniki, Greece.
{geokarag@ieee.org}

*Abstract*—The present paper is devoted to the evaluation of energy detection based spectrum sensing over different multipath fading and shadowing conditions. This is realized by means of a unified and versatile approach that is based on the particularly flexible mixture gamma distribution. To this end, novel analytic expressions are firstly derived for the probability of detection over MG fading channels for the conventional single-channel communication scenario. These expressions are subsequently employed in deriving closed-form expressions for the case of square-law combining and square-law selection diversity methods. The validity of the offered expressions is verified through comparisons with results from respective computer simulations. Furthermore, they are employed in analyzing the performance of energy detection over multipath fading, shadowing and composite fading conditions, which provides useful insighs on the performance and design of future cognitive radio based communication systems.

## I. INTRODUCTION

The need for efficient utilization of spectrum resources has become a fundamental requirement in modern wireless networks, mainly due to the aforementioned spectrum scarcity and the ever-increasing demand for higher data rate applications and Internet services [1]. In this context, cognitive radio (CR) communications is a particularly interesting wireless technology that has been proposed as an effective method that is capable of mitigating the spectrum scarcity by adapting their transmission parameters according to the respective environment [2]. To this end, cognitive radios have been shown to be highly efficient in maximizing spectrum utilization due to their inherent spectrum sensing capability. In a CR network environment, users are categorized in either primary users (PUs) or secondary users (SUs). Based on this, the former are the ones who have been typically assigned licensed spectrum slots, and hence, have higher priority, whereas the latter are accessing vacant frequency bands opportunistically.

Based on the above, numerous spectrum sensing techniques have been proposed over the past decade and can be classified into three main categories, namely, energy detection (ED), matched filter detection and cyclostationary or feature detection. One of the earliest methods is the likelihood ratio test (LRT) [3], which although it has been considered optimal, its technical exploitation is rather limited and impractical as it requires the exact knowledge of the signal-to-noise ratio (SNR) distributions as well as the corresponding channel information [4]. On the contrary, matched filter detection techniques [5], [6] typically require accurate synchronization and exact information about the transmitted signal waveform, such as its bandwidth and modulation type. Likewise, cyclostationary detection [7] uses the statistical properties of the transmitted signals to enhance the probability of detection. On the contrary, ED based sensing practically constitutes the most common detection method and has received considerable attention [8] [9] thanks to its low computational and implementation complexity. In ED, the presence of a PU signal is simply detected by comparing the output of the energy detector with a pre-determined energy threshold which depends on the *a priori* knowledge of the noise power level [10]. Therefore, poor knowledge of the noise power level leads to a high probability of false alarm and an SNR floor. Based on this, several analyses have been proposed for resolving this issue by estimating the noise power level, e.g. see [5], [11], [12] and the references therein. For instance, the authors in [12] proposed an iterative algorithm that optimizes the decision threshold for fulfilling the false alarm probability requirement. Unlike the conventional ED based spectrum sensing methods, which rely on the statistical covariance of the received signal, do not require the knowledge of the noise power level since their operation relies on the the difference that statistical covariance matrices of the received signal and the noise. See e.g. [13] and references therein.

It has been also extensively shown that fading phenomena create detrimental effects on the performance of conventional and emerging wireless communications, including cognitive radio systems. In this context, the ED performance over multipath fading channels, such as Rayleigh, Rician, and Nakagami-$m$ was analyzed in [14], and [15], respectively, whereas the corresponding performance over the more generalized $\kappa - \mu$ and $\kappa - \mu$ extreme fading channels was investigated in [16]. However, in addition to multipath fading, in most scenarios, the received signal is also degraded by shadowing effects since it has been shown that multipath and shadowing effects typically occur simultaneously [17]. Therefore, it is evident that there is an undoubted necessity to quantify and analyze the CR performance over composite multipath/shadowing fading channels [18]. Nevertheless, it has been shown that such an analysis is particularly tedious, since composite fading models can only be represented by cumbersome, if not intractable, infinite integrals. For example, the probability of detection of the ED based spectrum sensing over Nakagami-lognormal (NL) fading channels was addressed in [17]; yet, the offered solution is semi-analytic, as it is not represented in closed form, while the impact of fading and shadowing effects is evaluated numerically. Based on this, several alternative models that characterize the composite fading channels have been shown to provide simplified performance analysis for the CR networks. For example, in the analyses of [19]–[23], the $\mathcal{K}$ distribution is utilized to study the ED performance over Rayleigh/Lognormal (RL) channels. The energy detector performance for the MoG distribution [24] is derived in [25]. In [26], the sufficiency and optimality of cooperative wireless sensor networks that are based on energy detection is analyzed over NLOS fading environments, where zero-mean Gaussian mixtures are assumed as a viable model for NLOS fading channels. A unified and versatile analyses over the ED performance can be made feasible through the use of more recent generalized composite fading models, such as $\kappa - \mu$/Inverse-Gaussian [27] and $\eta - \mu$/Inverse-Gaussian [28] models.

In the present paper, we consider the generic and versatile mixture gamma (MG) based approach to derive new exact expressions for the average detection probability over generalized and composite fading channels. Specifically, a simple closed-form expression is derived for the case of integer values of the involved scale parameter $\beta_k$. It is recalled here that the MG model [29], [30] has been proposed as an alternative model to various generalized and composite fading channels, namely, Lognormal [30, eq. (6)], Weibull [30, eq. (5)], NL [30, eq. (9)], $\mathcal{K}_G$ [29, eq. (8)], $\eta - \mu$ [29, eq. (10)], $\kappa - \mu$ [29, eq. (17)], Hoyt [29, eq. (15)], and Rician [29, eq. (20)] channels. This model is both accurate and flexible to represent all aforementioned fading channels and thus, it constitutes a generic unified fading model. In the present analysis, the derived average detection probability is also extended to the case of diversity reception by means of the square-law combining (SLC) and square-law selection (SLS) schemes. It is noted that the probability of detection of the MG model has been derived in [28], yet, the solution provided therein is different from the approach in this paper. In addition, the present paper also provides novel and useful expressions for the SLC and SLS schemes.

The reminder of the paper is organized as follows: Section II provides a brief description of the system model, while Section III is devoted to the derivation of average detection probability expressions using the MG model with and without diversity reception. Analytical and numerical simulation results are presented in Section IV, while closing remarks are provided in Section V.

## II. SYSTEM MODEL

In a typical opportunistic cognitive radio configuration, a secondary user, which is assumed to employ energy detection for spectrum sensing, aims to determine whether a PU utilizes its assigned frequency band or not. Here, we assume that the channel gains, $h_j$, are independent and identically distributed (*i.i.d.*) and are modeled using the generalized MG distribution, where $j = 1, ..., L$. Furthermore, the received signal copies at the SU node and $j^{th}$ antenna can have two possible hypotheses, modeled as

$$\begin{aligned}\mathcal{H}_0 &: y_j(t) = v_j(t) \\ \mathcal{H}_1 &: y_j(t) = h_j(t) + v_j(t),\end{aligned} \quad (1)$$

where $\mathcal{H}_0$ and $\mathcal{H}_1$ represent the absence and presence of a signal, respectively, $s(t)$ corresponds to the transmitted signal from the PU, with energy $E_s = \mathbb{E}[|s(t)|^2]$, and $v_j(t) \sim \mathcal{CN}(0, \sigma_n^2)$ is the circularly symmetrical complex additive white Gaussian noise (AWGN). Here, the SU utilizes an energy detector that compares the amplitudes $|y_j|_{j=1}^M$ of the received signal to a threshold $\lambda$. Therefore, the output of this process for each antenna can be represented as follows

$$Z_j = |y_j|^2 \underset{\mathcal{H}_0}{\overset{\mathcal{H}_1}{\gtrless}} \lambda, \quad (2)$$

where the time index $t$ has been omitted for the sake of notational simplicity.

For the conventional case of AWGN channels, the conditional detection and false-alarm probabilities are determined with the aid of [31], namely

$$P_d = Q_u(\sqrt{2\gamma_j}, \sqrt{\lambda_n}), \quad (3)$$

$$P_f = \frac{\Gamma(u, \frac{\lambda_n}{2})}{\Gamma(u)}, \quad (4)$$

where $u$ is time-bandwidth product, $Q_u(\cdot, \cdot)$ is the generalized Marcum-Q function [32], $\Gamma(\cdot, \cdot)$ is the upper incomplete gamma function [33, eq. (8.35)], $\Gamma(\cdot)$ is the standard gamma function [33, eq. (8.31)], $\lambda_n = \lambda/\sigma_n^2$ is the normalized threshold, and $\gamma_j = \frac{|h_j|^2 E_s}{2\sigma_n^2}$ is the instantaneous SNR of the $j^{th}$ PU-SU link.

As the probability of false-alarm is based on the null hypothesis, it remains the same regardless of the involved fading conditions. Thus, in the subsequent sections, we focus on the derivation of the average detection probability for both the conventional single-channel communication and for multi-channel communications with diversity reception.

## III. PROBABILITY OF DETECTION OVER COMPOSITE FADING CHANNELS

### A. Single-antenna Scenario

It is recalled that the MG distribution is a generic and versatile distribution since it has been shown capable of providing accurate representation of several generalized and composite fading models. The corresponding probability density function (PDF) can be expressed as [21]

$$f_\gamma(x) = \sum_{k=1}^{C} \frac{\alpha_k}{\gamma_0} \left(\frac{x}{\gamma_0}\right)^{\beta_k-1} \exp\left(-\frac{\zeta_k x}{\gamma_0}\right), \quad (5)$$

where the scale and shape parameters of the $k^{th}$ component are denoted by $\beta_k$ and $\zeta_k$, respectively. Furthermore, the mixing coefficient of the $k^{th}$ component is denoted by $\alpha_k$, having the constraints, $0 \le \alpha_k\Gamma(\beta_k)/\zeta_k^{\beta_k} \le 1$ and $\sum_{k=1}^{C} \alpha_k\Gamma(\beta_k)/\zeta_k^{\beta_k} = 1$. To this effect, the average probability of detection for the MG distribution can be written as

$$\overline{P}_{d,MG} = \sum_{k=1}^{C} \frac{\alpha_k}{\gamma_0} \int_0^\infty Q_u(\sqrt{2x}, \sqrt{\lambda_n}) \cdot \left(\frac{x}{\gamma_0}\right)^{\beta_k-1} e^{-\frac{\zeta_k x}{\gamma_0}} dx. \quad (6)$$

Here, an exact closed-form expression is derived under the assumption that $\beta_k$ is a positive integer, i.e. $\beta_k \in \mathbb{N}$. This is realized with the aid of Theorem 1 in [34, eq. (3)] and by carrying out some long but basic algebraic simplifications yielding

$$\begin{aligned}\overline{P}_{d,MG} &= \sum_{k=1}^{C} \frac{\alpha_k \Gamma(\beta_k) \Gamma(u, \frac{\lambda_n}{2})}{\Gamma(u)\zeta_k^{\beta_k}} + \sum_{k=1}^{C}\sum_{l=0}^{\beta_k-1} \frac{\alpha_k \Gamma(\beta_k)}{\gamma_0^{\beta_k}} \\ &\quad \times \frac{\left(\frac{\lambda_n}{2}\right)^u {}_1\mathcal{F}_1(l+1, u+1, \frac{\lambda_n/2}{1+\frac{\zeta_k}{\gamma_0}})}{u!\left(\frac{\zeta_k}{\gamma_0}\right)^{\beta_k-l}\left(1+\frac{\zeta_k}{\gamma_0}\right)^{l+1} \exp(\frac{\lambda_n}{2})},\end{aligned} \quad (7)$$

where ${}_1\mathcal{F}_1(.,.,.)$ is the confluent hypergeometric function [33, eq. (9.210.1)]. It is noted here that the above expression has a relatively simple algebraic representation which renders it convenient to handle both analytically and numerically since the confluent hypergeometric function, ${}_1\mathcal{F}_1(.,.,.)$, is included as built-in function in popular software packages such as MATLAB, MAPLE and MATHEMATICA. It is worth noting that our derivations obtained in (7) coincides numerically with the expressions for the case of Rayleigh fading channel in [14] and [15].

### B. Diversity Reception

*1) Square-Law Combining:* Under SLC, the received signals from each branch are integrated, squared, and then summed up. It is also recalled that SLC is similar to the maximal-ratio combining scheme in the sense that the total instantaneous SNR at the output of the combiner is equivalent to that in MRC, i.e.

$$\gamma_\Sigma = \sum_{l=1}^{L} \gamma_l. \quad (8)$$

Nevertheless, SLC does not require channel estimation [34]. As a result, the conditional false-alarm probability would follow (4), with $u$ replaced by $Lu$. In order to evaluate the corresponding average detection probability, it is essential to derive the PDF of $\gamma_\Sigma$. To this end, for the case of $L=2$, the PDF of $\gamma_\Sigma$ can be obtained as follows:

$$\begin{aligned}f_{\gamma_\Sigma}^{(2)} &= \int_0^\gamma f_{\gamma_1}(x) f(\gamma-x) dx = \sum_{i=1}^{C}\sum_{j=1}^{C} \frac{\alpha_i \alpha_j}{\gamma_0^{\beta_i+\beta_j}} \\ &\quad \times \int_0^\gamma x^{\beta_i-1} e^{-\frac{\zeta_i}{\gamma_0}x}(\gamma-x)^{\beta_j-1} e^{-\frac{\zeta_j}{\gamma_0}(\gamma-x)} dx.\end{aligned} \quad (9)$$

In order to evaluate (9), we split the solution into two scenarios, namely when $\zeta_i = \zeta_j$ and $\zeta_i \ne \zeta_j$. In the former scenario, eq. (9) reduces to the following integral

$$\begin{aligned}f_{\gamma_\Sigma}^{(2)}|_{(\zeta_i=\zeta_j)} &= \sum_{i=1}^{C}\sum_{j=1}^{C} \frac{\alpha_i}{\gamma_0^{\beta_i}} \frac{\alpha_j}{\gamma_0^{\beta_j}} e^{-\frac{\zeta_j}{\gamma_0}\gamma} \\ &\quad \times \int_0^\gamma x^{\beta_i-1}(\gamma-x)^{\beta_j-1} dx.\end{aligned} \quad (10)$$

By performing the change of variables $u = \frac{x}{\gamma}$ and with the aid of [33, eq. (8.380)] and the functional relation in [33, eq. (8.384)], we obtain the following closed-form solution

$$f_{\gamma_\Sigma}^{(2)}|_{(\zeta_i=\zeta_j)} = \sum_{i=1}^{C}\sum_{j=1}^{C} \frac{\alpha_i \alpha_j}{\gamma_0^{\beta_i+\beta_j}} \frac{\Gamma(\beta_i)\Gamma(\beta_j)}{\Gamma(\beta_i+\beta_j)} e^{-\frac{\zeta_j}{\gamma_0}\gamma} \gamma^{\beta_i+\beta_j-1}. \quad (11)$$

On the contrary, for the case $\zeta_i \ne \zeta_j$, eq. (9) is solved with the aid of the binomial theorem in [33, eq. (1.111)] and under the assumption that $\beta_j \in \mathbb{N}$. To this effect, the representation in (9) can be equivalently re-written as follows

$$\begin{aligned}f_{\gamma_\Sigma}^{(2)}|_{(\zeta_i\ne\zeta_j)} &= \sum_{i=1}^{C}\sum_{j=1}^{C}\sum_{l=0}^{\beta_j-1} \binom{\beta_j-1}{l}(-1)^l \frac{\alpha_i \alpha_j}{\gamma_0^{\beta_i+\beta_j}} \\ &\quad \times \frac{\gamma^{\beta_j-l-1}}{e^{\frac{\zeta_j}{\gamma_0}\gamma}} \int_0^\gamma x^{\beta_i+l-1} e^{-\frac{x}{\gamma_0}(\zeta_i-\zeta_j)} dx.\end{aligned} \quad (12)$$

Evidently, the above integral can be expressed in closed-form with the aid of [33, eq. (8.350.1)] yielding

$$\begin{aligned}f_{\gamma_\Sigma}^{(2)}|_{(\zeta_i\ne\zeta_j)} &= \sum_{i=1}^{C}\sum_{j=1}^{C}\sum_{l=0}^{\beta_j-1} \binom{\beta_j-1}{l} \frac{(-1)^l \alpha_i \alpha_j}{\gamma_0^{\beta_j-l}(\zeta_i-\zeta_j)^{\beta_i+l}} \\ &\quad \times \gamma^{\beta_j-l-1} e^{-\frac{\zeta_j}{\gamma_0}\gamma} \gamma\left(\beta_i+l, \frac{\gamma(\zeta_i-\zeta_j)}{\gamma_0}\right),\end{aligned} \quad (13)$$

where $\gamma(a,x) \triangleq \int_0^x t^{a-1} e^{-t} dt$ denotes the lower incomplete gamma function. Thus, by expressing the $\gamma(a,x)$ function according to [33, eq. (9.352.6)], one obtains the following closed-form expression,

$$f_{\gamma_\Sigma}^{(2)}|_{(\zeta_i \neq \zeta_j)} = \sum_{i=1}^{C} \sum_{j=1}^{C} \sum_{l=0}^{\beta_j-1} \binom{\beta_j-1}{l} \frac{(-1)^l \alpha_i \alpha_j}{\gamma_0^{\beta_j-l}}$$
$$\times (\zeta_i - \zeta_j)^{-\beta_i-l} \gamma^{\beta_j-l-1} e^{-\frac{\zeta_j}{\gamma_0}\gamma} \Gamma(\beta_i+l)$$
$$\times (1 - e^{-\frac{\gamma(\zeta_i-\zeta_j)}{\gamma_0}} \sum_{t=0}^{\beta_i+l-1} \frac{\left(\frac{\gamma(\zeta_i-\zeta_j)}{\gamma_0}\right)^t}{t!}), \quad (14)$$

which is valid for $\beta_i \in \mathbb{N}$, while

$$f_{\gamma_\Sigma}^{(2)} = f_{\gamma_\Sigma}^{(2)}|_{(\zeta_i=\zeta_j)} + f_{\gamma_\Sigma}^{(2)}|_{(\zeta_i \neq \zeta_j)}. \quad (15)$$

By following the same methodology, a similar expression can be obtained for $f_{\gamma_\Sigma}^{(3)}$ as in (16) and (17) at the top of the next page.

It is noted here that the above methodology allows the derivation of similar expressions for $f_{\gamma_\Sigma}^{(4)}$, $f_{\gamma_\Sigma}^{(5)}$ and so forth.

Based on this, the corresponding average detection probability is readily obtained by

$$\overline{P}_{d,\Sigma}^{(L)} = \int_0^\infty Q_{Lu}(\sqrt{2\gamma_\Sigma}, \sqrt{\lambda}) f_{\gamma_\Sigma}^{(L)}(\gamma_\Sigma) d\gamma_\Sigma. \quad (18)$$

For the case of $L = 2$ and by inserting (11) and (14) in (18), it follows that

$$\overline{P}_{d,\Sigma}^{(2)}|_{(\zeta_i=\zeta_j)} = \sum_{i=1}^{C} \sum_{j=1}^{C} \frac{\alpha_i \alpha_j}{\gamma_0^{\beta_i+\beta_j}} \frac{\Gamma(\beta_i)\Gamma(\beta_j)}{\Gamma(\beta_i+\beta_j)}$$
$$\times \int_0^\infty \frac{Q_u(\sqrt{2\gamma_\Sigma},\sqrt{\lambda})e^{-\frac{\zeta_j}{\gamma_0}\gamma}}{\gamma^{-(\beta_i+\beta_j-1)}} d\gamma, \quad (19)$$

and

$$\overline{P}_{d,\Sigma}^{(2)}|_{(\zeta_i \neq \zeta_j)} = \sum_{i=1}^{C} \sum_{j=1}^{C} \sum_{l=0}^{\beta_j-1} \binom{\beta_j-1}{l} \frac{(-1)^l \alpha_i \alpha_j \Gamma(\beta_i+l)}{\gamma_0^{\beta_j-l}(\zeta_i-\zeta_j)^{\beta_i+l}}$$
$$\times I_1(\gamma_0) - \sum_{i=1}^{C} \sum_{j=1}^{C} \sum_{l=0}^{\beta_j-1} \sum_{t=0}^{\beta_i+l-1} \binom{\beta_j-1}{l}$$
$$\times \frac{(-1)^l \alpha_i \alpha_j \Gamma(\beta_i+l)}{t! \gamma_0^{\beta_j-l+t}(\zeta_i-\zeta_j)^{\beta_i+l-t}} I_2(\gamma_0), \quad (20)$$

where

$$I_1(\gamma_0) = \int_0^\infty \frac{Q_u(\sqrt{2\gamma_\Sigma},\sqrt{\lambda})\gamma^{\beta_j-l-1}}{e^{\frac{\zeta_j}{\gamma_0}\gamma}} d\gamma, \quad (21)$$

and

$$I_2(\gamma_0) = \int_0^\infty \frac{Q_u(\sqrt{2\gamma_\Sigma},\sqrt{\lambda})\gamma^{\beta_j+t-l-1}}{e^{\frac{\zeta_i}{\gamma_0}\gamma}} d\gamma. \quad (22)$$

Notably, the involved integrals in (19), (21), and (22) have the same algebraic representation as (6). Therefore, by utilizing Theorem 1 in [34, eq. (3)] and after some algebraic manipulations yields the closed-form expressions in (23) and (24), at the top of the next page.

In the same context, by following a similar methodology one can obtain the average detection probability for higher diversity orders while the probability of false alarm remains unchanged, i.e. $P_{f,\Sigma} = P_f$ in (6).

*2) Square-Law Selection:* Under SLS, the branch with the maximum $\gamma_j$ is selected as follows [14]

$$\gamma_{\text{SLS}} = \max_{j=1,..,L}(\gamma_j). \quad (25)$$

Under $\mathcal{H}_0$, the false-alarm probability for the SLS scheme can be expressed as

$$P_{f,\text{SLS}} = 1 - \Pr(\gamma_{\text{SLS}} < \lambda_n | \mathcal{H}_0). \quad (26)$$

Substituting (25) in (26), we obtain

$$P_{f,\text{SLS}} = 1 - \Pr(\max(\gamma_1, \gamma_2, .., \gamma_L) < \lambda_n | \mathcal{H}_0). \quad (27)$$

Accordingly, this translates to [35]

$$P_{f,\text{SLS}} = 1 - [1 - P_f]^L. \quad (28)$$

Similarly, the unconditional probability of detection over the AWGN channel is obtained by

$$P_{d,\text{SLS}} = 1 - \prod_{j=1}^{L} \left[1 - Q_u(\sqrt{2\gamma_j}, \sqrt{\lambda_n})\right]. \quad (29)$$

Hence, averaging (29) over (6) yields the unconditional probability of detection under the SLS scheme, $\bar{P}_{d,\text{SLS}}$, which is given by

$$\bar{P}_{d,\text{SLS}} = 1 - \prod_{j=1}^{L} [1 - P_{d,MG}]. \quad (30)$$

To the best of the authors' knowledge, the offered analytic results have not been previously reported in the open technical literature.

## IV. NUMERICAL RESULTS AND DISCUSSIONS

As already mentioned, the derived expressions are applicable to numerous generalized and composite fading channels, such as NL, RL, $\mathcal{K}$, $\mathcal{K}_G$, $\eta - \mu$, $\kappa - \mu$, Hoyt, and Rician channels. In this section, we present corresponding analytical and simulation results for the receiver operating characteristic (ROC) with and without diversity over certain fading scenarios. To this end, Fig. 1 depicts the analytical and simulated average missed-detection probability, $1 - P_d$, versus the false-alarm probability for different fading conditions with no diversity. It is clearly shown that the analytical and simulated curves are in tight agreement thanks to the arbitrarily accurate representation of the MG distribution. It is also shown that the presented $\eta - \mu$ scenario exhibits the best ROC, which is expected since it represents a rather light fading scenario with $\eta = 3.5$ and $\mu = 15$.

Fig. 2 depicts the analytical and simulated ROC curves over several scenarios of the composite NL fading channel with SLC diversity scheme with $L = 2$. As expected, changing the multipath severity parameter, $m$, has more prominent influence on the detection performance than changing the shadowing parameter, $\zeta$. For example, at $P_f = 0.48$, increasing $m$ from 3 to 4 resulted in the ratio,

$$f^{(3)}_{\gamma_\Sigma}|_{(\zeta_i=\zeta_j=\zeta_k)} = \sum_{i=1}^{C}\sum_{j=1}^{C} \frac{\alpha_i\alpha_j\alpha_k}{\gamma_0^{\beta_i+\beta_j+\beta_k}} \frac{\Gamma(\beta_i)\Gamma(\beta_j)\Gamma(\beta_k)}{\Gamma(\beta_i+\beta_j+\beta_k)} e^{-\frac{\zeta_k}{\gamma_0}\gamma} \gamma^{\beta_i+\beta_j+\beta_k-1}, \qquad (16)$$

$$\begin{aligned}
f^{(3)}_{\gamma_\Sigma}|_{(\zeta_i\neq\zeta_j\neq\zeta_k)} &= \sum_{i=1}^{C}\sum_{j=1}^{C}\sum_{k=1}^{C}\sum_{l=0}^{\beta_j-1}\sum_{r=0}^{\beta_k-1} \frac{\alpha_i\alpha_j\alpha_k}{\gamma_0^{\beta_k-r}} \binom{\beta_j-1}{l}\binom{\beta_k-1}{r} \\
&\quad\times \frac{(-1)^{\beta_j}\Gamma(l+\beta_i)}{(\zeta_i-\zeta_j)^{l+\beta_i}\zeta_k^{r+\beta_j-l}} \frac{\gamma^{\beta_k-r-1}}{e^{\frac{\gamma}{\gamma_0}(\zeta_k+\zeta_j)}} \gamma\left(b_j+r-l, -\frac{\zeta_k\gamma}{\gamma_0}\right) \\
&\quad - \sum_{i=0}^{C}\sum_{j=0}^{C}\sum_{k=0}^{C}\sum_{l=0}^{\beta_j-1}\sum_{t=0}^{\beta_i+l-1}\sum_{r=0}^{\beta_k-1} (-1)^{l+r}\frac{\alpha_i\alpha_j\alpha_k}{\gamma_0^{\beta_k-r}} \binom{\beta_j-1}{l}\binom{\beta_k-1}{r} \\
&\quad\times \frac{\Gamma(l+\beta_i)(\zeta_i-\zeta_j)^{t-l-\beta_i}}{t!(\zeta_i-\zeta_j-\zeta_k)^{r+t+\beta_j-l}} \gamma^{\beta_k-r-1} e^{-\frac{\gamma}{\gamma_0}(\zeta_k+\zeta_j)} \gamma\left(b_j+r+t-l, \frac{\gamma(\zeta_i-\zeta_j-\zeta_k)}{\gamma_0}\right).
\end{aligned} \qquad (17)$$

$$\overline{P}^{(2)}_{d,\Sigma}|_{(\zeta_i=\zeta_j)} = \sum_{i=1}^{C}\sum_{j=1}^{C} \alpha_i\alpha_j\Gamma(\beta_i)\Gamma(\beta_j)\left[\frac{\Gamma(u,\frac{\lambda}{2})}{\zeta_j^{\beta_i+\beta_j}\Gamma(u)} + \sum_{n=0}^{\beta_i+\beta_j-1} \frac{\gamma_0^{-n}(\frac{\lambda}{2})^u {}_1\mathcal{F}_1(n+1,u+1,\frac{\frac{\lambda}{2}}{1+\frac{\zeta_j}{\gamma_0}})}{u!(\zeta_j)^{\beta_i+\beta_j-n}(1+\frac{\zeta_j}{\gamma_0})^{n+1}\exp(\frac{\lambda}{2})}\right], \qquad (23)$$

$$\begin{aligned}
\overline{P}^{(2)}_{d,\Sigma}|_{(\zeta_i\neq\zeta_j)} &= \sum_{i=1}^{C}\sum_{j=1}^{C}\sum_{l=0}^{\beta_j-1} \binom{\beta_j-1}{l} \frac{(-1)^l\alpha_i\alpha_j\Gamma(\beta_i+l)}{\gamma_0^{\beta_j-l}(\zeta_i-\zeta_j)^{\beta_i+l}} \\
&\quad\times \left[\frac{\Gamma(\beta_j-l)\Gamma(u,\frac{\lambda}{2})}{(\frac{\zeta_j}{\gamma_0})^{\beta_j-l}\Gamma(u)} + \sum_{n=0}^{\beta_j-l-1} \frac{(\frac{\lambda}{2})^u\Gamma(\beta_j-l)}{u!(\frac{\zeta_j}{\gamma_0})^{\beta_j-l-n}} \frac{{}_1\mathcal{F}_1(n+1,u+1,\frac{\frac{\lambda}{2}}{1+\frac{\zeta_j}{\gamma_0}})}{(1+\frac{\zeta_j}{\gamma_0})^{n+1}\exp(\frac{\lambda}{2})}\right] \\
&\quad - \sum_{i=1}^{C}\sum_{j=1}^{C}\sum_{l=0}^{\beta_j-1}\sum_{t=0}^{\beta_i+l-1} \binom{\beta_j-1}{l} \frac{(-1)^l\alpha_i\alpha_j\Gamma(\beta_i+l)}{t!\gamma_0^{\beta_j-l+t}(\zeta_i-\zeta_j)^{\beta_i+l-t}} \\
&\quad\times \left[\frac{\Gamma(\beta_j+t-l)\Gamma(u,\frac{\lambda}{2})}{(\frac{\zeta_i}{\gamma_0})^{\beta_j+t-l}\Gamma(u)} + \sum_{n=0}^{\beta_j+t-l-1} \frac{(\frac{\lambda}{2})^u\Gamma(\beta_j+t-l) {}_1\mathcal{F}_1(n+1,u+1,\frac{\frac{\lambda}{2}}{1+\frac{\zeta_i}{\gamma_0}})}{u!(\frac{\zeta_i}{\gamma_0})^{\beta_j+t-l-n}(1+\frac{\zeta_i}{\gamma_0})^{n+1}\exp(\frac{\lambda}{2})}\right].
\end{aligned} \qquad (24)$$

$\frac{1-P_d|_{m=4}}{1-P_d|_{m=3}} = 4.47$, while reducing $\zeta$ from 4 to 1 improved the ROC curve by only $\frac{1-P_d|_{\zeta=1\,dB}}{1-P_d|_{\zeta=4\,dB}} = 1.40$. Fig. 3 depicts the analytical and simulated ROC curves over one scenario of the composite NL fading channel with SLS scheme with varying $L$. Comparing Figs. 2 and 3 exhibits that SLC performs better than SLS for $L = 2$; yet, the improvement is not significant for this particular case. For instance, at $P_{f,SLS} = P_{f,\Sigma} = 0.48$, the corresponding improvement ratio was $\frac{1-P_{d,\Sigma}}{1-P_{d,SLS}} = 1.80$. Also, one can observe how effective is the spatial diversity in combating the severity of the multipath fading and shadowing effects.

## V. CONCLUSIONS

We proposed a unified framework for the performance analysis of an energy detector in generalized and composite MG-based fading channels. Novel analytical expressions for the average detection probability have been derived for both the single-antenna case and the multiple-antenna case with square-law combining and square-law selection schemes. The derived expressions have been shown to be both generalized in terms of fading characterization and algebraically versatile.

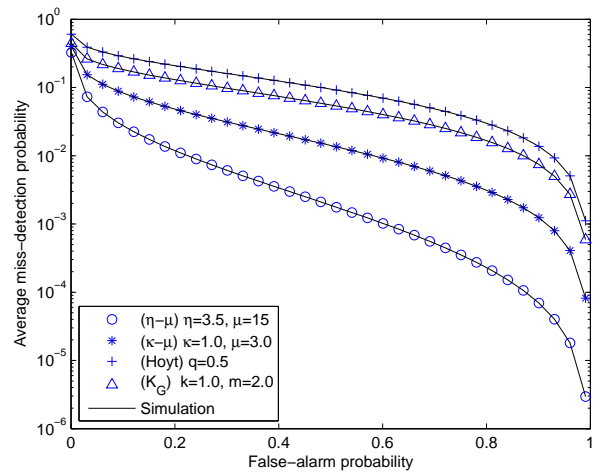

Figure 1. Complementary ROC for various fading channels and no diversity, with $\gamma_0 = 10\,\text{dB}$, $u = 2$.

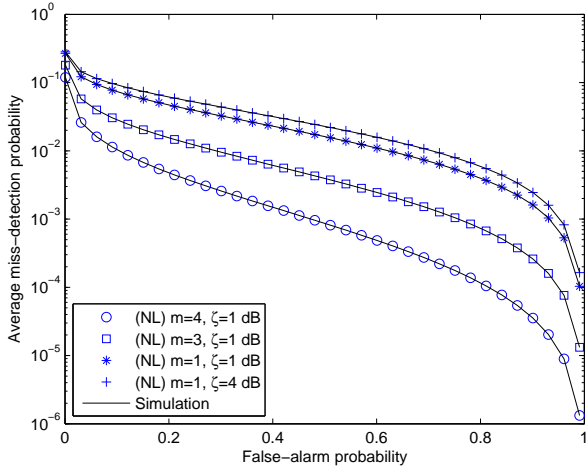

Figure 2. Complementary ROC for composite NL fading scenarios and SLC scheme, with $L = 2$, $\gamma_0 = 10\,\text{dB}$, $u = 2$.

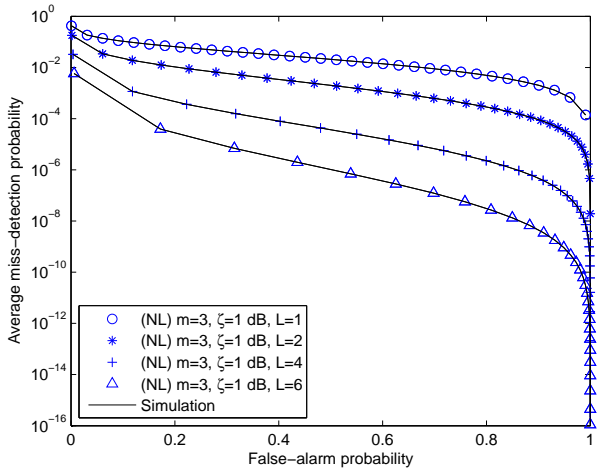

Figure 3. Complementary ROC for selected NL fading scenario and SLS scheme, with varying $L$ and $\gamma_0 = 10\,\text{dB}$, $u = 2$.